# $H^2$ Optimal Coordination of Homogeneous Agents Subject to Limited Information Exchange

Daria Madjidian, Leonid Mirkin, and Anders Rantzer

*Abstract*—Controllers with a diagonal-plus-low-rank structure constitute a scalable class of controllers for multi-agent systems. Previous research has shown that diagonal-plus-low-rank control laws appear as the optimal solution to a class of multi-agent $H^2$ coordination problems, which arise in the control of wind farms. In this paper we show that this result extends to the case where the information exchange between agents is subject to limitations. We also show that the computational effort required to obtain the optimal controller is independent of the number of agents and provide analytical expressions that quantify the usefulness of information exchange.

## I. INTRODUCTION

Large-scale systems are characterized by the presence of numerous subsystems (agents) having their own sensory and actuation abilities. Control of such systems, also known as distributed control, is being revived over the last decade due to networking and integration trends, efficiency demands, etcetera. A major challenge in distributed control design is to cope with constraints on information exchange between agents. Such constraints can be due to physical limitations (e.g., agents might only have access to local measurements) or they can be introduced artificially in order to reduce information processing and improve the implementational scalability of the control law.

In general, information exchange restrictions have adverse effects on the tractability of control design problems [1–3]. Yet these effects can be alleviated in situations when the information exchange topology "agrees" with the structure of the problem (the plant and control goals) [4–6]. It is thus important to understand what scalable information structure fits the considered application. The best studied in this context are sparsity-based information topologies (decentralized control), with non-zero elements corresponding to permitted information exchange between agents, see [7, 8] and the references therein. There are also applications, where interaction takes place through the *average behavior* of agents. For instance, in power systems generators must coordinate their power production to balance the total load on the network [9]. Hence, the power generated by a unit is directly coupled to the total net power imbalance.

Control structures based on aggregate (e.g., average) information are appealing from a large scale perspective due to its implementational scalability and low communication demands. Variations of this information structure have been studied in the context of the control of ensembles [10], biological systems [11], broadcast control [12], robust control [13], etcetera. Moreover, controllers with such information exchange mechanisms are shown to be optimal in $H^\infty$ problems over symmetrically interconnected systems [14] and in LQR coordination problems arising in the control of wind farms [15].

The result of [15] is the starting point of the current research, so we shall discuss it in more details. Specifically, a homogeneous group of autonomous agents that are coupled through a constraint on their average behavior was studied. It was shown that the optimal (centralized) solution consists of a diagonal (i.e., fully decentralized) term complemented by a rank-one component, which coordinates the agents based on their weighted-average state measurements. An additional scalability property attractive in large-scale applications is that the computational effort required to obtain the solution is independent of the number of agents.

In this paper we provide additional insight into the class of diagonal-plus-low-rank control laws by expanding the class of problems for which they are optimal. A potential limitation of the solution of [15] is that it assumes instantaneous information exchange between the agents. This might not be feasible in applications, where communication resources are limited. To account for these limitations, we modify the problem formulation by imposing additional constraints on the off-diagonal elements of the controller (Section II). Such constraints include time-delays, sampled-data processing, bandwidth limitations, etc. In Section III, we provide an abstract solution to the multi-agent problem in terms of the solution to a local, uniformly constrained control problem for a stand-alone agent. The main result is that the scalability properties, that were discussed above for the case of perfect information exchange in [15], extend to the case with limited information exchange as well. In particular, the optimal control law has a diagonal-plus-rank-one structure and can be obtained by solving a single local control problem. Based on established results in the literature, in Section IV we provide complete analytical solutions for two classes of communication constraints: delayed information exchange (§IV-A) and sampled-data information exchange (§IV-B). An illustrative example is presented then in Section V and concluding remarks, justifying a direction for the future research, are provided in Section VI.

*Notation:* The transpose of a matrix $M$ is denoted by $M'$. By $e_i$ we refer to the $i$th standard basis of an Euclidean space and by $I_n$ to the $n \times n$ identity matrix (we drop the dimension subscript when the context is clear). The lower linear fractional transformation [16, Sec. 10.1] is denoted as $\mathcal{F}_l(\cdot, \cdot)$. The notation $\otimes$ is used for the Kronecker product of matrices:

$$A \otimes B := \begin{bmatrix} a_{11}B & \cdots & a_{1m}B \\ \vdots & \ddots & \vdots \\ a_{p1}B & \cdots & a_{pm}B \end{bmatrix},$$

where $a_{ij}$ stands for the $(i, j)$ entry of $A$. In particular, $I_n \otimes M$ is a compact notation for the block-diagonal matrix having $n$ equal diagonal entries, $M$; $\sum_i (e_i e_i') \otimes M_i$ is the block-diagonal matrix with diagonal entries $M_i$; $\sum_i e_i \otimes M_i$ is the block-column matrix built of blocks $M_i$ with the same column dimension. The $H^2$-norm of a system $G$ is denoted $\|G\|_2$. The $H^2$ space and its norm are well-defined notions for both time-invariant [16, Sec. 4.3] and periodically time-varying [17, Sec. 9.1] linear systems.

This research was supported by the Swedish Research Council through the LCCC Linnaeus Center, the European Commission through the project AEOLUS, the Bernard M. Gordon Center for Systems Engineering at the Technion, and B. and G. Greenberg Research Fund (Ottawa).

D. Madjidian is with the Laboratory for Information and Decision Systems, Massachusetts Institute of Technology, Cambridge, MA 02139, USA. E-mail: dariam@mit.edu.

L. Mirkin is with the Faculty of Mechanical Engineering, Technion—IIT, Haifa 32000, Israel. E-mail: mirkin@technion.ac.il.

A. Rantzer is with the Department of Automatic Control, Lund University, Box 118, SE–221 00 Lund, Sweden. E-mail: anders.rantzer@control.lth.se.

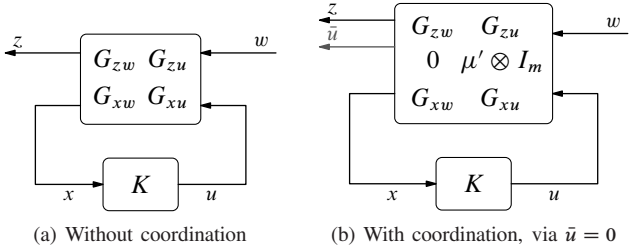

Fig. 1. Aggregate standard state-feedback $H^2$ problems

(a) Without coordination  (b) With coordination, via $\bar{u} = 0$

## II. PROBLEM FORMULATION

We study the problem of coordinating $\nu$ uncoupled homogeneous systems (agents), described by the following dynamics:

$$\dot{x}_i(t) = Ax_i(t) + B_w w_i(t) + B_u u_i(t) \quad (1)$$

where $x_i(t) \in \mathbb{R}^n$ are (measured) state vectors, $u_i(t) \in \mathbb{R}^m$ are control inputs, and $w_i(t) \in \mathbb{R}^n$ are exogenous disturbances. Associated with each agent is the local regulated variable

$$z_i(t) := C_z x_i(t) + D_{zu} u_i(t), \quad (2)$$

which reflects the local objectives of the agent.

Aggregating (1)–(2) for $i = 1, \ldots, \nu$, the local problems can be cast as the standard state-feedback $H^2$ problem [16, §14.8.1] depicted in Fig. 1(a). Here $w$, $z$, $u$, and $x$ are the aggregate disturbance, regulated output, control input, and measured state vector, respectively (e.g., $w := \sum_{i=1}^{\nu} e_i \otimes w_i$), and the generalized plant

$$\begin{bmatrix} G_{zw}(s) & G_{zu}(s) \\ G_{xw}(s) & G_{xu}(s) \end{bmatrix} = \left[\begin{array}{c|cc} I_\nu \otimes A & I_\nu \otimes B_w & I_\nu \otimes B_u \\ \hline I_\nu \otimes C_z & 0 & I_\nu \otimes D_{zu} \\ I_{\nu n} & 0 & 0 \end{array}\right].$$

Each sub-block of this generalized plant is block diagonal due to decoupled dynamics and objectives of the agents. Hence, if no other requirements were imposed, the optimal solution $K$ would be block diagonal as well.

Coordination among the agents is imposed by requiring that

$$\sum_{i=1}^{\nu} \mu_i u_i(t) =: \bar{u}(t) \equiv 0, \quad (3)$$

where $\mu_i$ may be viewed as a mass of the $i$th system. This constraint effectively requires the "center of mass" of all agents to behave as

$$\dot{\bar{x}}(t) = A\bar{x}(t) + B_w \bar{w}(t), \quad (4)$$

where $\bar{x} := \sum_i \mu_i x_i$ and $\bar{w} := \sum_i \mu_i w_i(t)$.

Complementing the setup in Fig. 1(a) with constraint (3), we end up with the setup depicted in Fig. 1(b), where $\mu := \begin{bmatrix} \mu_1 & \cdots & \mu_\nu \end{bmatrix}'$. The problem of minimizing the $H^2$ norm of the system $T_{zw}$ from $w$ to $z$, under the constraint that the system $T_{\bar{u}w}$ from $w$ to $\bar{u}$ is zero, is equivalent to the problem studied in [15]. It was show that the solution has the following scalable form

$$u_i(t) = F_\alpha(x_i(t) - \mu_i \bar{x}(t)), \quad (5)$$

where $F_\alpha$ is the LQR gain associated with the local, uncoordinated, problems. In this control law, the only information needed to coordinate the agents is the center of mass state, $\bar{x}$. This information, however, must be accessible instantaneously, which might not be feasible if communication resources are limited.

To account for potential communication limitations, in this paper we propose to introduce additional constraints upon the controller. Because inter-agent communication takes place through the off-diagonal elements of $K$, we constrain them to belong to a *subspace*, say $\mathcal{K}_c$, of the space of causal linear systems. In other words, we require that

$$K_{ij} \in \mathcal{K}_c, \quad \text{whenever } i \neq j \quad (6)$$

where the partitioning of $K$ is compatible with that of the signals $x$ and $u$. Several commonly considered communication constraints, such as time delays, sampling, and bandwidth limitations, may be expressed as in (6). Delays and sampling constraints will be addressed explicitly in Section IV.

The problem formulation considered in this paper is

$$\underset{\text{stabiliz. } K}{\text{minimize}} \quad \|T_{zw}\|_2 \quad (7a)$$
$$\text{subject to} \quad T_{\bar{u}w} = 0 \quad (7b)$$
$$K \text{ satisfies (6) for a given subspace } \mathcal{K}_c \quad (7c)$$

where $T_{zw}$ and $T_{\bar{u}w}$ are the closed-loop transfer functions in Fig. 1(b) from $w$ to $z$ and $\bar{u}$, respectively. We implicitly assume here that the $H^2$ norm is a well-defined notion for a given $\mathcal{K}_c$.

*Remark 2.1:* The coordination constraint (3) can be replaced by the more general requirement $\bar{u} = \bar{F}\bar{x}$, where $\bar{F}$ can be viewed as a "gain" shaping the "$A$" matrix of the center of mass in (4). However, this requirement can be reduced to (3) by a mere shift of the control variables as $u_i = v_i + \bar{F}x_i$. We therefore can consider the simpler version, (3), without any loss of generality. ▽

## III. ABSTRACT SOLUTION

In this section we solve (7) in a general form, without specifying a particular form of the constraint set $\mathcal{K}_c$. The only information about $\mathcal{K}_c$ that is required to formulate the solution is the assumption that it is a linear subspace. We also need to assume that

$\mathcal{A}_1$: $A$ is Hurwitz,

$\mathcal{A}_2$: $B_w$ is square and nonsingular,

$\mathcal{A}_3$: $\mu'\mu = 1$ and none of the entries of $\mu$ is zero.

Assumption $\mathcal{A}_1$ is necessary for the stabilizability of the overall system because the dynamics of the center of mass (4) are not affected by the control signal. $\mathcal{A}_2$ effectively says that the null space of $G_{xw}$ is trivial. It is made to avoid technical issues related to uniqueness of the corresponding optimal $H^2$ solution (see [16, §14.8.1]). The normalization part in $\mathcal{A}_3$ is introduced to simplify the exposition and can be relaxed. Finally, if $\mu_i = 0$, then the $i$th system is not a part of the coordination problem and can therefore be excluded from the analysis.

The solution of (7) is based on the solution to the $H^2$ state-feedback problem associated with the generalized plant

$$G_\alpha(s) = \begin{bmatrix} G_{\alpha 11}(s) & G_{\alpha 12}(s) \\ G_{\alpha 21}(s) & G_{\alpha 22}(s) \end{bmatrix} = \left[\begin{array}{c|cc} A & B_w & B_u \\ \hline C_z & 0 & D_{zu} \\ I & 0 & 0 \end{array}\right]$$

and a controller $K_\alpha$. The problem is formulated as follows:

$$\underset{\text{stabiliz. } K_\alpha}{\text{minimize}} \quad \|\mathcal{F}_l(G_\alpha, K_\alpha)\|_2 \quad (8a)$$
$$\text{subject to} \quad K_\alpha \in \mathcal{K}_c \quad (8b)$$

For various $\mathcal{K}_c$ of interest, problem (8) can be solved by available techniques (see Section IV for two particular cases). Meanwhile, we do not elaborate on these solutions. What we need is to assume that

$\mathcal{A}_4$: problem (8) is well posed,

in the sense that its optimal solution $K_{\alpha,\text{opt}}$ exists and is unique. The resulting optimal performance is $\gamma_\alpha := \|\mathcal{F}_l(G_\alpha, K_{\alpha,\text{opt}})\|_2$. We also need the quantity

$$\gamma_0 := \|G_{\alpha 11}\|_2 = \sqrt{\text{tr}(B_w' X B_w)} \geq \gamma_\alpha,$$

where $\bar{X} \geq 0$ is the solution to the Lyapunov equation
$$A'\bar{X} + \bar{X}A + C_z'C_z = 0.$$

The solution to (7), which is the main result of the paper, is given in the following theorem, whose proof is presented in §III-A:

*Theorem 3.1:* Let $\mathcal{A}_{1\text{-}4}$ hold true. Then the optimal achievable
$$\|T_{zw}\|_2^2 = (\nu - 1)\gamma_\alpha^2 + \gamma_0^2 \tag{9}$$
and it is attained by the control law
$$u_i = K_{\alpha,\text{opt}}(x_i - \mu_i \bar{x}), \tag{10}$$
where $\bar{x} = \sum_i \mu_i x_i$. ▽

A noteworthy outcome of Theorem 3.1 is that the two scalability properties of the solution of [15], which studied a version of (7) without communication constraints (7b), extend to the case when these constraints are added. First, we only need to solve the local uncoordinated problem (8) to form the optimal control law in Theorem 3.1. In other words, the computational effort is independent of the number of agents $\nu$. Second, although the optimal control law (15) is not decentralized (due to the presence of $\bar{x}$), the only global computation needed to form it is a single (scaled) averaging operation, exactly as in (5).

### A. Proof of Theorem 3.1

We start with a technical result, which reduces the (unorthodox) constraint (6) on the off-diagonal parts of the controller to a uniform constraint on the whole $K$.

*Lemma 3.2:* Let $\mathcal{A}_{2,3}$ hold. Then (7c) is satisfied together with (7b) only if the whole $K \in \mathcal{K}_c$.

*Proof:* $\mathcal{A}_2$ implies that $G_{xw}(s) = I \otimes ((sI - A)^{-1} B_w)$ is square and nonsingular. Because $G_{xu}$ is strictly causal, the loop in Fig. 1(b) is well posed and we have that
$$T_{\bar{u}w} = (\mu' \otimes I)K(I - G_{xu}K)^{-1}G_{xw} = 0 \iff (\mu' \otimes I)K = 0,$$
which involves only the controller. Thus, (7b) holds iff
$$\sum_{i=1}^\nu \mu_i K_{ij} = 0 \iff K_{jj} = -\sum_{i \neq j} \frac{\mu_i}{\mu_j} K_{ij},$$
for all $j = 1, \ldots, \nu$. Because $\mathcal{K}_c$ is a subspace, the latter equality combined with (6) implies that $K_{jj} \in \mathcal{K}_c$ as well. ■

*Remark 3.1:* Assumption $\mathcal{A}_2$ can in principle be relaxed here. It can be shown that if it does not hold, the admissible diagonal elements of $K$ that are not in $\mathcal{K}_c$ do not affect $T_{zw}$ anyway. But the solution is not unique then. ▽

Having reduced (7) to a problem with uniformly constrained $K$, we may apply the technique used in [15] to decouple the coordination constraint (7b). Namely, let $U \in \mathbb{R}^{\nu \times \nu}$ be a unitary matrix such that $U\mu = e_1$, i.e. $U'$ comprises the left singular vectors of $\mu$. Define new state vector $\tilde{x} := (U \otimes I_n)x$, control input $\tilde{u} := (U \otimes I_m)u$, exogenous input $\tilde{w} := (U \otimes I_n)w$, and regulated signal $\tilde{z} := (U \otimes I_p)z$. The relation between these "tilded" signals is the same as the relation between their originals, which can be verified using the equality $(U \otimes I_{n_1})(I_\nu \otimes M) = (I_\nu \otimes M)(U \otimes I_{n_2})$ holding for every $M \in \mathbb{R}^{n_1 \times n_2}$. For example, the transfer function from $\tilde{w}$ to $\tilde{z}$ is
$$G_{\tilde{z}\tilde{w}}(s) = (U \otimes I_p)G_{zw}(s)(U' \otimes I_n)$$
$$= (U \otimes I_p)(I_\nu \otimes (C_z(sI - A)^{-1}B_w))(U' \otimes I_n)$$
$$= (UU') \otimes (C_z(sI - A)^{-1}B_w) = G_{zw}(s).$$

Taking into account that $(\mu' \otimes I)u = (e_1' \otimes I)\tilde{u}$, the system in Fig. 1(b) can then be equivalently presented as shown in Fig. 2, where
$$\tilde{K} := (U \otimes I_m) K (U' \otimes I_n).$$

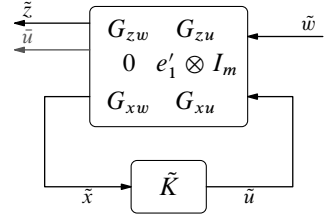

Fig. 2. Decoupled problem in transformed coordinates

Because $\tilde{K}$ is a linear invertible function of $K$ and $\mathcal{K}_c$ is a subspace, $K \in \mathcal{K}_c$ iff $\tilde{K} \in \mathcal{K}_c$ and because $U$ is unitary, $\|T_{\tilde{z}\tilde{w}}\|_2 = \|T_{zw}\|_2$ for every $K$. Thus, (7) can be solved via solving the $H^2$ problem associated with the system in Fig. 2.

The advantage of the latter is that it is decoupled. Indeed, the coordination constraint in terms of $\tilde{u}$ reads $\bar{u} = \tilde{u}_1$, so that (7b) prespecifies the first component of $\tilde{u}$ and has no effect on the others. Therefore, the $H^2$ problem for the setup in Fig. 2 splits into $\nu$ independent problems, the first of which is solved by the zero controller and the others are $\nu - 1$ copies of (8). Consequently, the optimal $\tilde{K}_{\text{opt}} = (I_\nu - e_1 e_1') \otimes K_{\alpha,\text{opt}}$, from which
$$K_{\text{opt}} = (I_\nu - \mu \mu') \otimes K_{\alpha,\text{opt}}. \tag{11}$$

This controller produces the control law (10). The optimal cost is then in the form (9) and this completes the proof. ■

### B. Cost distribution among agents

In this subsection we study how the overall optimal performance is distributed among the agents and how different components of $w$ affect local regulated variables $z_i$. To this end, consider the closed-loop systems $T_{z_i w_j}$ from the $j$th exogenous input $w_j$ to $z_i$ under the optimal controller (11). It is readily verified that
$$K_{\text{opt}}(I - G_{xu}K_{\text{opt}})^{-1} = (I - \mu\mu') \otimes \big(K_{\alpha,\text{opt}}(I - G_{\alpha 22}K_{\alpha,\text{opt}})^{-1}\big),$$
from which
$$T_{zw} = I \otimes G_{\alpha 11} + (I \otimes G_{\alpha 12})K_{\text{opt}}\big(I - G_{xu}K_{\text{opt}}\big)^{-1}(I \otimes G_{\alpha 21})$$
$$= I \otimes T_{\alpha,\text{opt}} + (\mu\mu') \otimes (G_{\alpha 11} - T_{\alpha,\text{opt}}),$$
where $T_{\alpha,\text{opt}} := \mathcal{F}_l(G_\alpha, K_{\alpha,\text{opt}})$. Hence, $T_{z_i w_j} = (e_i' \otimes I)T_{zw}(e_j \otimes I)$ can be expressed as
$$T_{z_i w_j} = \begin{cases} T_{\alpha,\text{opt}} + \mu_i^2(G_{\alpha 11} - T_{\alpha,\text{opt}}) & \text{if } j = i \\ \mu_i \mu_j (G_{\alpha 11} - T_{\alpha,\text{opt}}) & \text{otherwise} \end{cases}$$

The following result says that under a mild technical assumption the $H^2$ norm of $T_{z_i w_j}$ is a function of $\gamma_0 = \|G_{\alpha 11}\|_2$ and the optimal performance $\gamma_\alpha$ of (8):

*Proposition 3.3:* Let $K_{\alpha,\text{opt}} G_{\alpha 22} K_{\alpha,\text{opt}} \in \mathcal{K}_c$. Then
$$\|T_{z_i w_j}\|_2^2 = \gamma_\alpha^2 \delta_{ij} + \mu_i^2 \mu_j^2 (\gamma_0^2 - \gamma_\alpha^2),$$
where $\delta_{ij}$ is the Kronecker delta.

*Proof:* Let $T_\alpha := \mathcal{F}_l(G_\alpha, K_\alpha)$. Differentiating it with respect to $K_\alpha$ yields
$$dT_\alpha = G_{\alpha 12}(I - K_\alpha G_{\alpha 22})^{-1} dK_\alpha (I - G_{\alpha 22} K_\alpha)^{-1} G_{\alpha 21}.$$

It follows from $\mathcal{A}_4$ that $T_{\alpha,\text{opt}} \perp \mathcal{S}$, where
$$\mathcal{S} := G_{\alpha 12}(I - K_{\alpha,\text{opt}} G_{\alpha 22})^{-1} \mathcal{K}_c (I - G_{\alpha 22} K_{\alpha,\text{opt}})^{-1} G_{\alpha 21}.$$

To see this, split $T_{\alpha,\text{opt}} = T_\mathcal{S} + T_{\mathcal{S}\perp}$, where $T_\mathcal{S}$ and $T_{\mathcal{S}\perp}$ are the orthogonal projections of $T_{\alpha,\text{opt}}$ on $\mathcal{S}$ and its orthogonal complement, respectively. In particular, $T_\mathcal{S} = G_{\alpha 12}(I - K_{\alpha,\text{opt}} G_{\alpha 22})^{-1} K_\mathcal{S} (I - G_{\alpha 22} K_{\alpha,\text{opt}})^{-1} G_{\alpha 21}$ for some $K_\mathcal{S} \in \mathcal{K}_c$. Let $d\zeta$ be an infinitesimal positive step. Then, the choice $K_\alpha = K_{\alpha,\text{opt}} - K_S \text{d}\zeta$ results in $T_\alpha = T_{\alpha,\text{opt}} - T_S \text{d}\zeta$, so that $\|T_\alpha\|^2 = (1-\text{d}\zeta)^2\|T_S\|^2 + \|T_{S^\perp}\|^2 \le \|T_{\alpha,\text{opt}}\|^2$ with the equality attainable iff $T_S = 0$.

Next, set $K_\alpha = K_{\alpha,\text{opt}} - K_{\alpha,\text{opt}} G_{\alpha 22} K_{\alpha,\text{opt}}$. Then $K_\alpha \in \mathcal{K}_c$, and

$$T_{\alpha,\text{opt}} \perp G_{\alpha 12}(I - K_{\alpha,\text{opt}} G_{\alpha 22})^{-1} K_\alpha (I - G_{\alpha 22} K_{\alpha,\text{opt}})^{-1} G_{\alpha 21}$$
$$\perp G_{\alpha 12} K_{\alpha,\text{opt}}(I - G_{\alpha 22} K_{\alpha,\text{opt}})^{-1} G_{\alpha 21} = T_{\alpha,\text{opt}} - G_{\alpha 11},$$

and consequently, by the Pythagorean theorem,

$$\|G_{\alpha 11} - T_{\alpha,\text{opt}}\|_2^2 = \|G_{\alpha 11}\|_2^2 - \|T_{\alpha,\text{opt}}\|_2^2 = \gamma_0^2 - \gamma_\alpha^2,$$

from which the result follows immediately. ■

*Remark 3.2:* The assumption in Proposition 3.3 can be expected to hold whenever $\mathcal{K}_c$ is a uniform constraint in (8). In particular, this is true for the two examples considered in Section IV. In general, the assumption is weaker than the quadratic invariance condition [5], which requires $K_\alpha G_{\alpha 22} K_\alpha \in \mathcal{K}_c$ for *all* $K_\alpha \in \mathcal{K}_c$. ▽

The overall cost of the $i$th agent, which is the $H^2$ norm of $T_{z_i w} := [\,T_{z_i w_1}\ \cdots\ T_{z_i w_\nu}\,]$, is an immediate corollary of Proposition 3.3:

*Corollary 3.4:* If the condition of Proposition 3.3 holds, then

$$\|T_{z_i w}\|_2^2 = \mu_i^2 \gamma_0^2 + (1 - \mu_i^2)\gamma_\alpha^2.$$

This cost can be interpreted from two points of view. First, the quantity $\gamma_0$ can be thought of as the optimal cost of (7) in the absence of information exchange between subsystems. Indeed, in this case (7b) must be satisfied *by each agent*, resulting in the optimal law $u_i = 0$. From this viewpoint, the quantity

$$\gamma_{\text{BoC}} := \gamma_0^2 - \|T_{z_i w}\|_2^2 = (1 - \mu_i^2)(\gamma_0^2 - \gamma_\alpha^2) \ge 0$$

characterizes the *benefit of cooperation* for the $i$th agent. If the number of agents increases, the (normalized) $\mu_i$'s normally decrease [15, §III-C.3] and coordination becomes more beneficial per agent.

Another way to look at $\|T_{z_i w}\|_2$ in Corollary 3.4 is to compare it with the performance of the $i$th agent attainable via solving (7a) without the coordination constraint (7b). No coordination is required in this case, so that the optimal controller is block-diagonal and the coordination constraint (6) is void. The optimal performance of each agent is then $\gamma_{\text{opt}} = \sqrt{\text{tr}(B_w' X_\alpha B_w)} \le \gamma_\alpha$, where $X_\alpha$ is the stabilizing solution of the corresponding Riccati equation (in fact, of (12) defined in the next section). The addition of the coordination requirement (7b) naturally leads to a performance deterioration. The quantity

$$\gamma_{\text{CoC}} := \|T_{z_i w}\|_2^2 - \gamma_{\text{opt}}^2 = \mu_i^2(\gamma_0^2 - \gamma_{\text{opt}}^2) + (1 - \mu_i^2)(\gamma_\alpha^2 - \gamma_{\text{opt}}^2)$$

can then be interpreted as the *cost of satisfying the coordination constraint* (7b). The first term in the right-hand side of this expression is exactly the cost of coordination in the absence of communication constraints (6), see [15, Prop. 3.2]. This term normally vanishes as the number of agents $\nu \to \infty$. The second term in the expression for $\gamma_{\text{CoC}}$ quantifies the deterioration of the local cost due to communication constraints. This term actually grows with the decrease of $\mu_i$. Since $\gamma_{\text{CoC}} = \mu_i^2(\gamma_0^2 - \gamma_\alpha^2) + (\gamma_\alpha^2 - \gamma_{\text{opt}}^2)$, it decreases with the increase of the number of coordinating agent, although does not vanish as in the case when no communication constraints are imposed.

## IV. Particular Cases of Communication Constraints

### A. Delayed information exchange

Communication limitations can be accounted for by artificially introducing a sufficiently large time-delay, say $h > 0$, into the communication channels. In terms of (6), this corresponds to

$$\mathcal{K}_c = \mathcal{K}_{\text{del}} := \{K : K(s) = e^{-sh} K_h(s) \text{ for a causal } K_h\}$$

for some $h > 0$, which is a linear subspace and is quadratically invariant with respect to any causal linear plant.

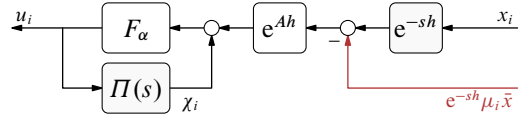

Fig. 3. A realization of the optimal controller for the $i$th system

Problem (8) in this case is a well-understood $H^2$ problem with single loop delay, which can be solved by available methods[1] [18, 19]. In this case $\mathcal{A}_4$ can be replaced with

$\mathcal{A}_5$: $\begin{bmatrix} A - j\omega I & B_u \\ C_z & D_{zu} \end{bmatrix}$ has full column rank $\forall \omega \in \mathbb{R}$,

$\mathcal{A}_6$: $D_{zu}' D_{zu} = I$.

Indeed, given $\mathcal{A}_1$, assumption $\mathcal{A}_5$ is necessary for the well-posedness of the unconstrained local problems. The normalization assumption $\mathcal{A}_6$ is introduced to simplify the exposition and can be relaxed to $D_{zu}' D_{zu} > 0$. If $\mathcal{A}_{1,5,6}$ hold true, then the algebraic Riccati equation

$$A' X_\alpha + X_\alpha A + C_z' C_z - (X_\alpha B_u + C_z' D_{zu})(B_u' X_\alpha + D_{zu}' C_z) = 0, \quad (12)$$

has a unique stabilizing solution $0 \le X_\alpha \le \bar{X}$ and the solution to (8) is

$$K_{\alpha,\text{opt}}(s) = (I + F_\alpha \Pi(s))^{-1} F_\alpha e^{Ah} e^{-sh}, \quad (13)$$

where $F_\alpha := -(B_u' X_\alpha + D_{zu}' C_z)$ and

$$\Pi(s) := \int_0^h e^{-(sI-A)\theta} \text{d}\theta B_u \quad (14a)$$
$$= (sI - A)^{-1}(B_u - e^{Ah} B_u e^{-sh}) \quad (14b)$$

is a stable FIR (finite impulse response) dead-time compensator. The optimal attainable $H^2$ performance with this controller is

$$\gamma_\alpha^2 = \text{tr}\left(B_w'\left(e^{A'h} X_\alpha e^{Ah} + \int_0^h e^{A't} C_z' C_z e^{At} \text{d}\theta\right) B_w\right)$$
$$= \text{tr}(B_w' \bar{X} B_w) - \text{tr}(B_w' e^{A'h}(\bar{X} - X_\alpha) e^{Ah} B_w).$$

Theorem 3.1 yields then that the optimal controller solving (7a) is in the form presented in Fig. 3. This control law can be also described by the following equation in the time domain:

$$u_i(t) = F_\alpha \hat{x}_i(t) - \mu_i F_\alpha e^{Ah} \bar{x}(t - h), \quad (15)$$

where

$$\hat{x}_i(t) := e^{Ah} x_i(t - h) + \int_{t-h}^{t} e^{A(t-\theta)} B_u u_i(\theta) \text{d}\theta \quad (16)$$

is the *mean-squared prediction* of $x_i(t)$ in (1) based on $x_i(\tau)$, $\tau \le t - h$, cf. [20, Lesson 16].

*Remark 4.1:* Although distributed-delay systems, like $\Pi$ in (14a), can be safely implemented, see [21] and the references therein, their implementation might be numerically involved. The implementation in our case, however, is simplified because the matrix $A$ is Hurwitz (by $\mathcal{A}_1$). Indeed, in this case $\Pi$ can be implemented in the equivalent form (14b), whose singularities at the eigenvalues of $A$ are removable. This transfer function can be implemented as

$$\dot{\chi}_i(t) = A \chi_i(t) + B_u u_i(t) - e^{Ah} B_u u_i(t - h), \quad (17)$$

which is a combination of a (stable) finite-dimensional system and a pure delay element, whose implementations are standard. Although this implementation involves pole-zero cancellations of all eigenvalues of $A$, the cancellations are *stable*. Hence, the implementation via (17) is internally stable and thus admissible. ▽

---

[1]Although these references study output-feedback versions of the problem, their adjustment to the "state feedback" case is fairly straightforward.

## B. Sampled-data information exchange

Another possibility to reduce the burden of communication, apart from using delays, is to exchange information only at some sampling instances. This leads to the following set:

$$\mathcal{K}_c = \mathcal{K}_{sd} := \{K : K = \mathcal{H}_h K_d \mathcal{S}_h \text{ for a discrete causal } K_d\},$$

where $\mathcal{S}_h$ and $\mathcal{H}_h$ are sampling (A/D) and hold (D/A) devices, respectively, which are assumed to be synchronized and with a sampling period $h > 0$. Hereafter, we assume that $\mathcal{S}_h$ is the ideal sampler transforming analog signals to discrete sequences as

$$x_d = \mathcal{S}_h x \iff x_d[k] = x(kh), \quad \forall k \in \mathbb{Z},$$

and $\mathcal{H}_h$ is the zero-order hold transforming discrete sequences to piecewise constant analog signals as

$$u = \mathcal{H}_h u_d \iff u(kh + \tau) = u_d[k] \quad \forall k \in \mathbb{Z}, \tau \in (0, h].$$

The set $\mathcal{K}_{sd}$ is a subspace and quadratically invariant with respect to any causal system.

The sampled-data version of (8) can be viewed as a particular case (state feedback) of the standard sampled-data $H^2$ problem extensively studied in the literature, see [22] and the references therein. The well-posedness assumption $\mathcal{A}_4$ can be replaced by

$\mathcal{A}_7$: $\begin{bmatrix} A & B_u \\ C_z & D_{zu} \end{bmatrix}$ has full column rank,

which, together with $\mathcal{A}_1$, guarantees that the sampled-data problem associated with (1) and (2) is non-singular [22, Cor. 5.2 (ii)]. If these two conditions hold, the discrete ARE

$$\hat{X}_\alpha = \hat{A}' \hat{X}_\alpha \hat{A} + \hat{Q} \\ - (\hat{A}' \hat{X}_\alpha \hat{B}_u + \hat{S})(\hat{B}'_u \hat{X}_\alpha \hat{B}_u + \hat{R})^{-1} (\hat{B}'_u \hat{X}_\alpha \hat{A} + \hat{S}'). \quad (18)$$

has a stabilizing solution $\hat{X}_\alpha \geq 0$, where

$$\begin{bmatrix} \hat{A} & \hat{B}_u \end{bmatrix} := \begin{bmatrix} e^{Ah} & \int_0^h e^{At} dt B_u \end{bmatrix} = \begin{bmatrix} I & 0 \end{bmatrix} \exp\left(\begin{bmatrix} A & B_u \\ 0 & 0 \end{bmatrix} h\right)$$

and

$$\begin{bmatrix} \hat{Q} & \hat{S} \\ \hat{S}' & \hat{R} \end{bmatrix} := \int_0^h \exp\left(\begin{bmatrix} A' & 0 \\ B'_u & 0 \end{bmatrix} t\right) \begin{bmatrix} C'_z \\ D'_{zu} \end{bmatrix} \\ \times \begin{bmatrix} C_z & D_{zu} \end{bmatrix} \exp\left(\begin{bmatrix} A & B_u \\ 0 & 0 \end{bmatrix} t\right) dt.$$

The optimal performance level in (8) is then

$$\gamma_\alpha^2 = \frac{1}{h} \text{tr}\left(B'_w \int_0^h \left(\int_0^t e^{A'\theta} C'_z C_z e^{A\theta} d\theta + e^{A't} \hat{X}_\alpha e^{At}\right) dt B_w\right) \\ = \frac{1}{h} \text{tr}\left(B'_w \int_0^h \left(\bar{X} - e^{A't} \bar{X} e^{At} + e^{A't} \hat{X}_\alpha e^{At}\right) dt B_w\right) \\ = \text{tr}(B'_w \bar{X} B_w) - \frac{1}{h} \text{tr}\left(B'_w \int_0^h e^{A't} (\bar{X} - \hat{X}_\alpha) e^{At} dt B_w\right)$$

and it is attained by the static control law

$$K_{\alpha,\text{opt}} = \mathcal{H}_h \hat{F}_\alpha \mathcal{S}_h.$$

where

$$\hat{F}_\alpha := -(\hat{B}'_u \hat{X}_\alpha \hat{B}_u + \hat{R})^{-1} (\hat{B}'_u \hat{X}_\alpha \hat{A} + \hat{S}').$$

Theorem 3.1 yields the control law

$$u_i(kh + \tau) = \hat{F}_\alpha (x_i(kh) - \mu_i \bar{x}(kh)) \quad (19)$$

at every $k = 0, 1, \ldots$ and $\tau \in (0, h]$.

*Remark 4.2:* One may think of several alterations of the subspace $\mathcal{K}_{sd}$. For example, the waveform of the control signal, i.e., the D/A part of the controller, may be considered a part of the design. Because the D/A part is implemented only locally, this alterations does not affect the inter-agent communication. A version of (8) in which the hold device is a part of the design was solved in [23]. The solution assumes $\mathcal{A}_{5,6}$ to guarantee that assumption $\mathcal{A}_4$ holds and results in the control law

$$u_i(kh + \tau) = F_\alpha e^{(A + B_u F_\alpha)\tau} (x_i(kh) - \mu_i \bar{x}(kh)), \quad (20)$$

where $F_\alpha$ is the continuous-time state feedback gain, the same as that appearing in (13), and the optimal performance level

$$\gamma_\alpha^2 = \text{tr}(B'_w \bar{X} B_w) - \frac{1}{h} \text{tr}\left(B'_w \int_0^h e^{A't} (\bar{X} - X_\alpha) e^{At} dt B_w\right),$$

where $X_\alpha$ is the stabilizing solution of (12). Performance of (20) is better than that of (19), while communication demands are the same.

Another potential modification of $\mathcal{K}_{sd}$ is to combine sampled-data and delay constraints. Problem (8) can then be solved by the approach of [24], both in the case of the zero-order and the optimal holds. ∇

## V. ILLUSTRATIVE EXAMPLE

Consider a formation of homogeneous vehicles described by

$$p_i = \frac{1}{s^2} (\tau_i + w_i), \quad (21)$$

where, $p_i$ is the position of the $i$th vehicles, $\tau_i$ is its thrust, and $w_i$ is a disturbance. The objectives are twofold:
1) the formation center of mass, defined as $\bar{p} := \frac{1}{\nu} \sum_{i=1}^\nu p_i$, follows a reference trajectory $\bar{r}(t)$ with bounded two first derivatives,
2) each vehicle tracks a fixed position relative to the center of mass, $r_i := \bar{p} + \delta_i$ for given constants $\delta_i \neq \delta_j$ and such that $\sum_i \delta_i = 0$.

We assume that each vehicle has perfect measurements of $p_i$ and $\dot{p}_i$ and that it knows $\delta_i$ and $\bar{r}$ and its first two derivatives, but that communication between vehicles is subject to sampled data constraints with the sampling period $h$, as described in §IV-B.

We start with the first objective. It is readily seen that the center of mass verifies $\bar{p} = (\bar{\tau} + \bar{w})/s^2$ with $\bar{\tau} := \frac{1}{\nu} \sum_i \tau_i$ and $\bar{w} := \frac{1}{\nu} \sum_i w_i$. The 2DOF state-feedback control law for this system,

$$\bar{\tau} = s^2 \bar{r} - (\kappa_1 s + \kappa_0) \bar{\epsilon} \quad (22)$$

with $\bar{\epsilon} := \bar{p} - \bar{r}$, renders the closed-loop error system

$$\bar{\epsilon} = \frac{1}{s^2 + \kappa_1 s + \kappa_0} \bar{w}, \quad (23)$$

independent of $\bar{r}$. By an appropriate choice of the gains $\kappa_0 > 0$ and $\kappa_1 > 0$ we can affect the disturbance sensitivity of the error behavior.

Because $\bar{\tau}$ is the average of the local thrusts $\tau_i$, the implementation of (22) requires coordination between the vehicles. It is readily seen that all $\tau_i$ that realize (22) can be parametrized as

$$\tau_i = (s^2 + \kappa_1 s + \kappa_0) \bar{r} - (\kappa_1 s + \kappa_0) p_i + \kappa_0 \delta_i + u_i, \quad (24)$$

where $u_i$ is an arbitrary signal satisfying $\sum_i u_i = 0$ (remember the assumption $\sum_i \delta_i = 0$). The term $\kappa_0 \delta_i$ is added to (24) to render $u_i = 0$ in the case when the perfect tracking conditions $\bar{p} = \bar{r}$ and $p_i = r_i$ are met. Substituting $\tau_i$ from (24) into (21), we end up with stable agents, in terms of deviation variables, of the form

$$y_i := p_i - \bar{r} - \delta_i = \frac{1}{s^2 + \kappa_1 s + \kappa_0} (u_i + w_i). \quad (25)$$

These systems correspond to (1) with

$$x_i = \begin{bmatrix} y_i \\ \dot{y}_i \end{bmatrix} \quad \text{and} \quad \begin{bmatrix} A | B_w | B_u \end{bmatrix} = \begin{bmatrix} 0 & 1 & 0 & 0 \\ -\kappa_0 & -\kappa_1 & 1 & 1 \end{bmatrix}$$

although $B_w$ above does not satisfy $\mathcal{A}_2$, the optimal solution to the corresponding $H^2$ problem is still unique, see [16, Prop. 14.9]).

Having set the average dynamics of the platoon, we use the remaining degrees of freedom to minimize the $H^2$ cost function based on the regulated signals
$$\tilde{z}_i = \begin{bmatrix} \sqrt{q_1}(p_i - r_i) \\ \sqrt{q_2}(\dot{p}_i - \dot{r}_i) \\ \tau_i - \ddot{r} \end{bmatrix} = \begin{bmatrix} \sqrt{q_1}(y_i - \bar{\epsilon}) \\ \sqrt{q_2}(\dot{y}_i - \dot{\bar{\epsilon}}) \\ u_i - \kappa_1 \dot{y}_i - \kappa_0 y_i \end{bmatrix}$$
for some weights $q_1 \geq 0$ and $q_2 \geq 0$. The term $\tau_i - \ddot{r}$ penalizes the deviation of $\tau_i$ from the "ideal" thrust, which meets both our objectives under zero disturbances. As in any realistic situation $\ddot{r} = 0$ in steady state, this term can be regarded as a penalty on the $i$th thrust.

Define now the unit vector $\mu := \sum_i e_i / \sqrt{\nu}$ and the aggregate output $y := \sum_i e_i y_i$. It can be verified that
$$\bar{\epsilon} = \frac{1}{\nu} \sum_i y_i = \frac{1}{\sqrt{\nu}} \mu' y \implies \sum_i e_i(y_i - \bar{\epsilon}) = (I_\nu - \mu\mu')y$$
Because $(I_\nu - \mu\mu')^2 = I_\nu - \mu\mu'$, we have that
$$\sum_i (y_i - \bar{\epsilon})^2 = y'(I_\nu - \mu\mu')y = \sum_i y_i^2 - \nu \bar{\epsilon}^2.$$
The last term here, $\bar{\epsilon}^2$, does not depend on $u_i$ as long as (3) holds. Hence, the optimization problem with the regulated outputs $\tilde{z}_i$ is equivalent, modulo a shift in the attainable performance, to that with the regulated variable
$$z_i = \begin{bmatrix} \sqrt{q_1} y_i \\ \sqrt{q_2} \dot{y}_i \\ u_i - \kappa_1 \dot{y}_i - \kappa_0 y_i \end{bmatrix} = \begin{bmatrix} \sqrt{q_1} & 0 \\ 0 & \sqrt{q_2} \\ -\kappa_0 & -\kappa_1 \end{bmatrix} x_i + \begin{bmatrix} 0 \\ 0 \\ 1 \end{bmatrix} u_i,$$
which is in form (2). In other words, the second objective can be cast as problem (7) with $\mathcal{K}_c = \mathcal{K}_{sd}$. This will result in $u_i$ acting as (19) for some calculated $\hat{F}_\alpha = [\begin{array}{cc} f_{\alpha 1} & f_{\alpha 2} \end{array}]$. The control law (24) then reads (using the fact that $y_i - \mu_i \bar{y} = p_i - \bar{p} - \delta_i$)
$$\tau_i(t) = \ddot{r}(t) + (\kappa_0 - f_{\alpha 1})\delta_i - \kappa_0(p_i(t) - \bar{r}(t)) - \kappa_1(\dot{p}_i(t) - \dot{\bar{r}}(t)) \\ + f_{\alpha 1}(p_i(kh) - \bar{p}(kh)) + f_{\alpha 2}(\dot{p}_i(kh) - \dot{\bar{p}}(kh)) \quad (26)$$
for $t \in [kh, (k+1)h)$, which uses sampled global and analog local measurements indeed. By an appropriate choice of the weights $q_1$ and $q_2$ we may then tune the behavior of the individual cars, say to strike a trade-off between maintaining a rigid formation and reducing the energy consumption. Our main point in this section, however, is to show how a more sophisticated problem can be handled within the proposed framework, so simulation results are not presented here.

## VI. CONCLUDING REMARKS

We have studied a large-scale state-feedback $H^2$ problem, in which a homogeneous group of autonomous agents is coupled through a constraint on their average behavior and where the information exchange between the agents is limited. It has been shown that for a range of communication restrictions, which includes time-delays, sampled-data processing, and bandwidth limitations, the problem can be reduced to an $H^2$ problem of the same dimension as that of a single agent. Moreover, the optimal controller for the original large-scale problem is composed of a diagonal (decentralized) term complemented by a rank-one coordination component. This structure, as well as the computational scalability of the solution, are the same as in the case without communication restrictions studied in [15].

A key step in proving our main result was to show that the communication limitations in combination with the hard constraint on the agents to coordinate their behavior, prevent each agent from using its full set of available information. This property sets a fundamental limitation on the achievable performance. In particular, unlike the case with perfect information exchange, the cost of coordination per agent does not vanish as the number of agents grows. A natural question is how this performance limitation changes when the coordination constraint is replaced with a *coordination incentive* (soft constraints). It is also of interest to understand if the diagonal-plus-rank-one structure and the computational scalability carry over to this case.